\documentclass[aps,prl,twocolumn,amsmath,amssymb]{revtex4}
\usepackage{graphicx}
\usepackage{color}
\usepackage{graphicx}
\usepackage{dcolumn}
\usepackage{bm}
\usepackage{amsmath}
\usepackage{amsfonts}
\usepackage{bbm}
\usepackage{subfigure}
\usepackage{setspace}
\newcommand{\beq}{\begin{eqnarray}}
\newcommand{\eeq}{\end{eqnarray}}

\newcommand{\bmp}{\noindent\begin{minipage}{16cm}}
\newcommand{\emp}{\end{minipage}\vskip 7mm} 

\def\drawbox#1#2{\hrule height#2pt
        \hbox{\vrule width#2pt height#1pt \kern#1pt
              \vrule width#2pt}
              \hrule height#2pt}

\def\Asym#1#2{\vcenter{\vbox{\drawbox{#1}{#2}
              \kern-#2pt 
              \drawbox{#1}{#2}}}}

\begin{document}
\title{\Large  WW Scattering in  Walking Technicolor
}
\author{Roshan {\sc Foadi}}
\email{roshan@fysik.sdu.dk}
\author{Francesco {\sc Sannino}}
\email{sannino@fysik.sdu.dk}
\affiliation{University of Southern Denmark, Campusvej 55, DK-5230 Odense M, Denmark.}


\begin{abstract}
We analyze the $WW$ scattering in scenarios of dynamical electroweak symmetry breaking of walking technicolor type. We show that in these theories there are regions of the parameters space allowed by the electroweak precision data, in which unitarity violation is delayed at tree level up to around 3-4 TeV without the inclusion of any sub-TeV resonances. \end{abstract}

\maketitle

The simplest argument often used to predict the existence of yet undiscovered particles at the TeV scale comes from unitarity of longitudinal gauge boson scattering amplitudes. If the electroweak symmetry breaking sector (EWSB) is weakly interacting, unitarity implies that new particle states must show up below 1 TeV, being these spin-0 isosinglets (the Higgs boson) or spin-1 isotriplets (e.g. Kaluza-Klein modes). A strongly interacting EWSB sector can however change this picture, because of the strong coupling between the pions (eaten by the longitudinal components of the standard model gauge bosons) and the other bound states of the strongly interacting sector. An illuminating example comes from QCD. In Ref. \cite{Harada:2003em}  it was shown that for six colors or more, the 770 GeV $\rho$ meson  is enough to delay the onset of unitarity violation of the pion-pion scattering amplitude up to well beyond 1 GeV. Here the 't Hooft large N limit was used, however an even lower number of colors is needed to reach a similar delay of unitarity violation when an alternative large N limit is used \cite{Sannino:2007yp}. Scaling up to the electroweak scale, this translates in a 1.5 TeV techni-$\rho$ being able to delay unitarity violation of longitudinal gauge boson scattering amplitudes up to 4 TeV or more. Such a particle would be harder to be discovered at the LHC and ILC. Such a model, however, would not be realistic for other reasons: a large contribution to the $S$ parameter \cite{Peskin:1990zt}, and large flavor changing neutral currents (FCNC) if the ordinary fermions acquire mass via an old fashioned extended technicolor sector (ETC), to mention the most relevant ones.

Walking Technicolor (WT) \cite{Holdom:1981rm,Eichten:1979ah,Lane:1989ej}
provides a natural framework to address these problems. In fact walking dynamics helps suppressing FCNC without preventing ETC from yielding realistic fermion masses. Notice that it is always possible to resort to new scalars to give mass to the ordinary fermions while having technicolor only in the gauge sector, or even marry technicolor with supersymmetry \cite{Simmons:1988pu,Simmons:1988fu,Dine:1990jd,Kagan:1990az,Kagan:1991gh,Carone:1992rh,Carone:1993xc,Gudnason:2006mk}. Furthermore certain WT models are in agreement with the constraints imposed by electroweak precision data (EWPD)  \cite{Foadi:2007se,Foadi:2007ue,Dietrich:2005jn}, since the walking dynamics itself naturally lowers the contribution to the $S$ parameter relative to a running theory \cite{Appelquist:1998xf}. Besides, new leptonic sectors \cite{Dietrich:2005jn}, which may be needed to avoid possible Witten topological anomalies, can render the overall $S$ parameter negative. The contributions from these sectors, not gauged under the technicolor gauge group, are calculable to any order in perturbation theory. A relevant question to ask is whether a walking regime can be achieved with a sufficiently small number of fermions. In the context of SU($N_{\rm TC}$) gauge theories with fermions in the fundamental representation, a large number $N_{\rm TF}$ of techniflavors is required to achieve walking dynamics, even for small values of $N_{\rm TC}$. If all technifermions have electroweak charge, this results in a large contribution to $S$ and other electroweak parameters. Recently, however, a new class of SU($N_{\rm TC}$) gauge theories with fermions in higher-dimensional representations have been argued to display near-to-conformal behavior already for small values of $N_{\rm TF}$ and $N_{\rm TC}$ \cite{Sannino:2004qp}. A complete catalogue, based on the Schwinger-Dyson approximation, of all possible $SU(N_{\rm TC})$ walking technicolor gauge theories with fermions in a given but arbitrary representation can be found in \cite{Dietrich:2006cm}. An all order $\beta$-function for any nonsupersymmetric, strongly interacting, and asymptotically free gauge theory has been suggested in \cite{Ryttov:2007cx}, thereby  allowing new methods for the investigation of nonsupersymmetric conformal windows.  The first hints of walking or possibly conformal dynamics for gauge theories with fermions in higher dimensional representations has appeared in \cite{Catterall:2007yx}, and a first benchmark for a working model of WT has been constructed in \cite{Foadi:2007ue}. Finally it should be mention that in this framework unification of the SM gauge couplings can be achieved \cite{Gudnason:2006mk}. 

An alternative scenario in which the $S$ parameter is naturally suppressed is provided by Custodial Technicolor (CT) \cite{Foadi:2007se}, in which an enhanced global symmetry \cite{Appelquist:1999dq} prohibits interactions between the spin-0 and the spin-1 sectors of the theory, thereby protecting all of the elctroweak parameters against large corrections. This feauture, however, renders the vector mesons totally helpless when it comes to unitarize the pion-pion scattery amplitudes. Therefore, a Higgs boson with mass around or below 1 TeV is always needed in CT. Based on these premises, it is interesting to analyze the pion-pion scattering in generic models of WT.

In order to extract predictions in presence of a strongly interacting sector and an asymptotically free gauge theory, we make use of the time-honored Weinberg sum rules (WSR) \cite{Weinberg:1967kj}, which are statements about the vector-vector minus axial-axial vacuum polarization functions, known to be sensitive to chiral symmetry breaking. Assuming a low energy spectrum consisting of a narrow vector-vector resonance and a narrow axial-vector resonance, the first WSR reads
\begin{eqnarray}
F_{\rm V}^2-F_{\rm A}^2=F_\pi^2 \ ,
\label{WSR1}
\end{eqnarray} 
where $F_{\rm V}$ ($F_{\rm A}$) is the decay constant of the vectorial (axial) resonance, and $F_\pi$ is the pion decay constant. In technicolor models $F_\pi=1/\sqrt2 G_F$ = 246 GeV. The second WSR, unlike the first one, receives important contributions from throughout the near conformal region, and reads \cite{Appelquist:1998xf,Foadi:2007ue}
\begin{eqnarray}
F_{\rm V}^2 M_{\rm V}^2-F_{\rm A}^2 M_{\rm A}^2=a\frac{8\pi^2}{d({\rm R})}F_\pi^4 \ , 
\label{WSR2}
\end{eqnarray}
where $d({\rm R})$ is the dimension of the representation of the underlying fermions, and $a$ is expected to be positive and ${\cal O}(1)$. In case of running dynamics we obtain $a=0$, and the second WSR recovers its familiar form. The parameter a is a non-universal quantity depending on the details of the underlying gauge theory expected to be positive and {\cal O}(1).
  Any other approach trying to model the walking behavior will have to reduce to ours. The fact that $a$ is positive and of order one in walking dynamics it is supported, indirectly, also from the work of Kurachi and Shrock \cite{Kurachi:2006ej}. At the onset of conformal dynamics the axial and the vector will be degenerate, i.e. $M_A=M_V=M$, using the first sum rule one finds via the second sum rule $a = d({\rm R})M^2/(8\pi^2 F^2_{\pi})$ leading to a numerical value of about 4- 5 from the approximate results in \cite{Kurachi:2006ej}.  We will however use only the constraints coming from the generalized WSRs  expecting them to be less model dependent.

There is also a ``zeroth'' sum rule, which is nothing but the definition of the $S$ parameter:
\begin{eqnarray}
S=4\pi\left[\frac{F_{\rm V}^2}{M_{\rm V}^2}-\frac{F_{\rm A}^2}{M_{\rm A}^2}\right] \ .
\label{WSR0}
\end{eqnarray}
Using Eq.~(\ref{WSR1}) and Eq.~(\ref{WSR2}) this becomes
\begin{eqnarray}
S=4\pi F_\pi^2 \left[\frac{1}{M_{\rm V}^2}+\frac{1}{M_{\rm A}^2}-a\frac{8\pi^2 F_\pi^2}{d({\rm R}) M_{\rm A}^2 M_{\rm V}^2}\right] \ .
\end{eqnarray}
The first two terms of this equation are the ordinary QCD-like contributions. The third term is negative and of the same order of the first two. Therefore, if the $\rho$ and $a_1$ mesons of QCD were resonances of a walking theory, the corresponding $S$ parameter would be considerably lower than the QCD value $S\simeq 0.3$. As mentioned before, this shows that the walking dynamics is not only important in suppressing FCNC, but also in lowering the contribution to the $S$ parameter relative to a theory with a running coupling constant. Since QCD's $S$ is approximately twice its perturbative estimate, it appears safe to estimate $S$ for any WT theory to be within $\pm$100\% the perturbative value, calculated from techniquark loops.
 Recent computations further support the reduction of $S$ in walking theories \cite{Kurachi:2007at}.
\begin{figure}
\begin{center}
\includegraphics[height=5cm,width=8cm]{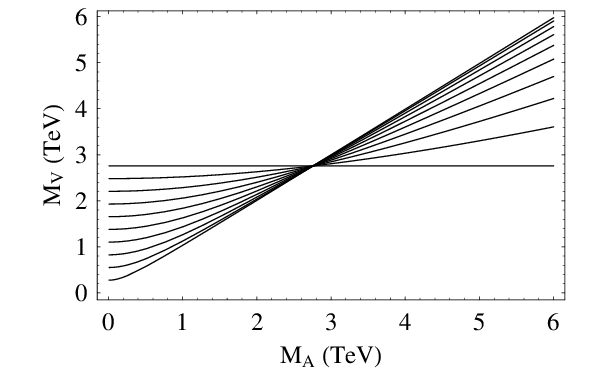}
\caption{$M_{\rm V}$ as a function of $M_{\rm A}$ for $S=0.1$, and different values of $g$. The curve with the lowest slope corresponds to the lowest value of $g$ compatible with the requirement $\Gamma_{\rm V}/M_{\rm V}\leq 0.5$. The curve with the largest slope corresponds to $g=\sqrt{8\pi/S}$: values of $g$ above this bound give $F_{\rm A}^2<0$, and must therefore be rejected. See below for the allowed region in the $(M_{\rm A},g)$ plane.}
\label{spectrum}
\end{center}
\end{figure}

\begin{figure*}
\begin{center}
\begin{tabular}{cc}
{\includegraphics[height=4.5cm,width=7cm]{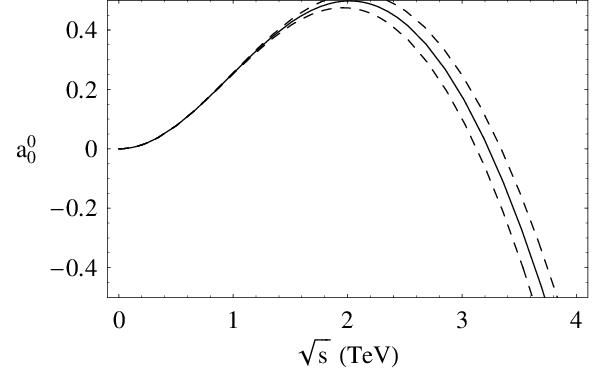}} ~~~&~~~ {\includegraphics[height=4.5cm,width=7cm]{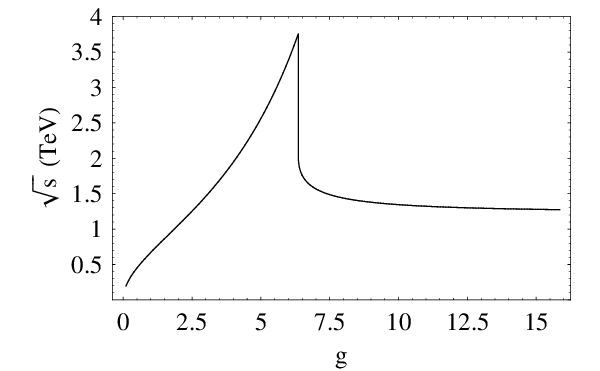}} \\ 
~~~{(a)} &~~~ {(b)}
\end{tabular}
\caption{(a) I=0, J=0 partial wave amplitude for $M_{\rm A}$=1.5 TeV, $S=0.1$, and three different values of $g$. The central curve corresponds to the largest delay of unitarity violation, up to $\sqrt{s}\simeq$ 3.7 TeV. The lowest curve corresponds to a slightly lower value of $g$, and violates unitarity just below 3.7 TeV. The upper curve corresponds to a slightly larger value of $g$, but it violates unitarity at a much lower energy, $\sqrt{s}\simeq$ 1.9 TeV. As a consequence there is a discontinuity in the plot of unitarity violation energy as a function of $g$ (b).}
\label{MA1TeV}
\end{center}
\end{figure*}
\begin{figure*}
\begin{center}
\begin{tabular}{cc}
{\includegraphics[height=4.5cm,width=7.5cm]{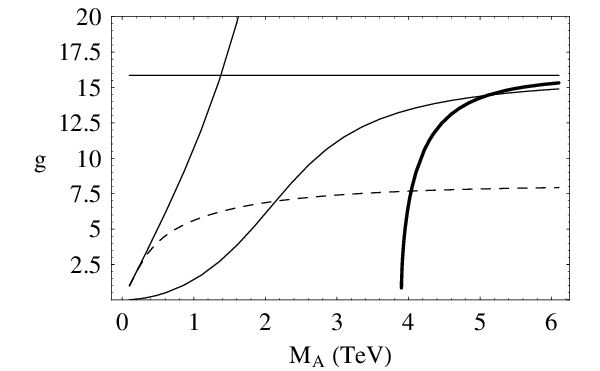}} &
{\includegraphics[height=4.5cm,width=7.5cm]{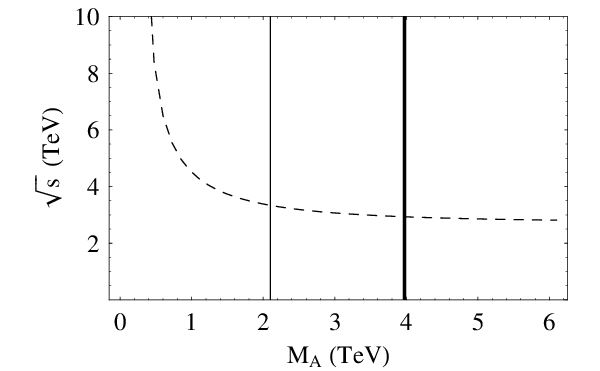}} \\
{\includegraphics[height=4.5cm,width=7.5cm]{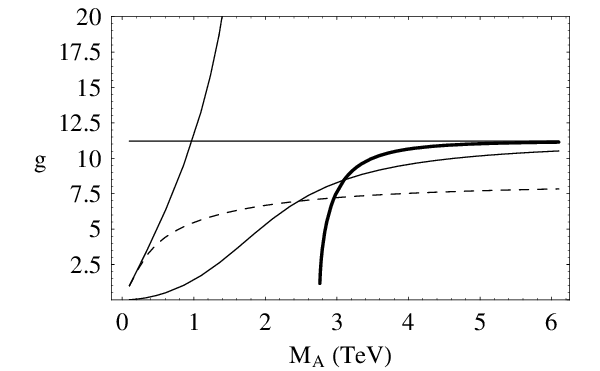}} &
{\includegraphics[height=4.5cm,width=7.5cm]{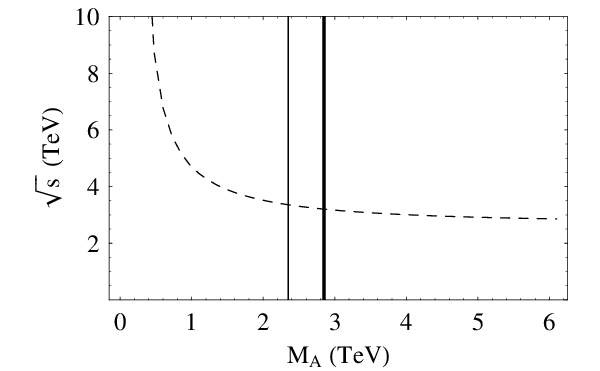}} \\
{\includegraphics[height=4.5cm,width=7.5cm]{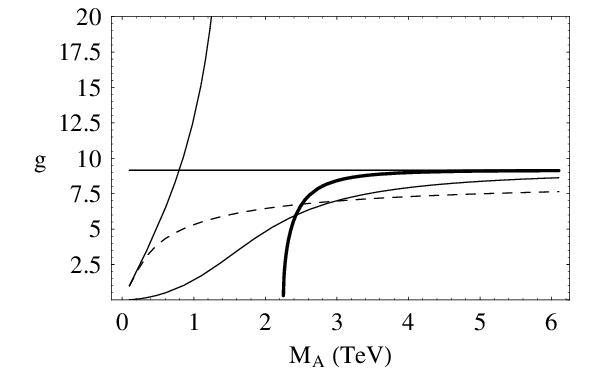}} &
{\includegraphics[height=4.5cm,width=7.5cm]{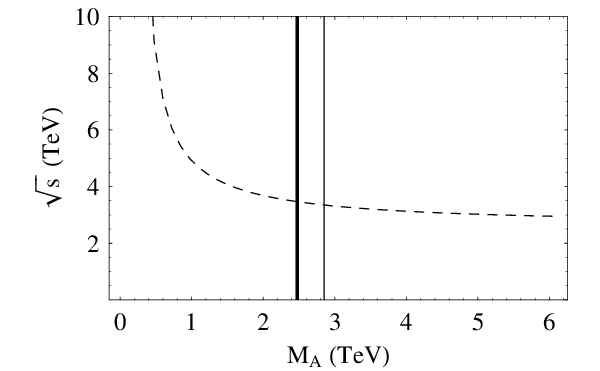}} \\
\ \ \ \ \ \ \ \ \ \ \ \ \  {(a)} & \ \ \ \ \ \ \ \ \ \ \ \ \ {(b)}
\end{tabular}
\caption{From top to bottom these figures correspond to $S=0.1$, $S=0.2$, and $S=0.3$. (a) The dashed line corresponds to $g_{\rm c}$ as a function of $M_{\rm A}$. The thin solid lines give bounds on $g$ and $M_{\rm A}$ coming from self consistency (no imaginary numbers) and the requirement that the vector-vector meson is a narrow state, as explained in the text. The thick solid line corresponds to $a=0$: along this curve both WSR's are satisfied in a running regime, while below the curve $a<0$, and the corresponding values of $g$ and $M_{\rm A}$ must be rejected. This curve is hit by $g_{\rm c}$, with a narrow width,  for $S=0.3$, but not for $S\leq 0.2$, proving that QCD-like theories are only unitarized, at the tree-level, by vector mesons for large values of $S$ or, which is the same, a large number of colors. (b) Unitarity violation scale along $g=g_{\rm c}$. To the right of the vertical thin line $\Gamma_{\rm V}/M_{\rm V}$ becomes greater than 0.5. To the right of the thick verical line $a$ becomes negative.}
\label{unit}
\end{center}
\end{figure*}

As a phenomelogically interesting example we consider minimal walking technicolor (MWT) \cite{Foadi:2007ue,Dietrich:2005jn,Sannino:2004qp}. This is an SU(2) gauge theory with two fermions in the adjoint representation. If these form an electroweak doublet, the perturbative contribution to the $S$ parameter will be $S=d(R)/6\pi=1/2\pi\simeq 0.16$. MWT is arguably the WT model with the lowest $S$. This value can be further reduced by adding to the theory new leptons (which in MWT are required to cure the Witten anomaly), with weak SU(2) mass splitting. With these ingredients the full estimates for $S$ and $T$ were shown to be within 1$\sigma$ of the experimental expectation values for a wide portion of the parameter space  \cite{Foadi:2007ue,Dietrich:2005jn,Sannino:2004qp}. This is true also for the more stringest tests based on the precise LEP parameters of Barbieri {\em et. al.} \cite{Barbieri:2004qk} explored for WT and CT in \cite{Foadi:2007se}. Taking $S\in(0.1,0.3)$ as a realistic estimate in MWT, and using Eq.~(\ref{WSR1}) and Eq.~(\ref{WSR0}), allows us to take $F_{\rm V}$ and $M_{\rm A}$ (for example) as the only independent inputs, since $F_\pi$ is known. In alternative to $F_{\rm V}$ we can take the ``gauge coupling'' $g$ of an effective model \cite{Foadi:2007ue} in which the vector mesons are treated as gauge fields. $F_{\rm V}$ is given by
\begin{eqnarray}
F_{\rm V}^2=\frac{2 M_{\rm V}^2}{g^2} \ .
\label{FV}
\end{eqnarray}
From Eq.~(\ref{WSR1}), Eq.~(\ref{WSR0}), and Eq.~(\ref{FV}) we find
\begin{eqnarray}
M_{\rm V}^2 &=& \left(1-\frac{g^2 S}{8\pi}\right)M_{\rm A}^2+\frac{g^2 F_\pi^2}{2} \label{MV} \\
F_{\rm V}^2 &=& \left(1-\frac{g^2 S}{8\pi}\right)\frac{2M_{\rm A}^2}{g^2} + F_\pi^2 \\
F_{\rm A}^2 &=& \left(1-\frac{g^2 S}{8\pi}\right)\frac{2M_{\rm A}^2}{g^2} \label{FA2} \ .
\end{eqnarray}
Eq.~(\ref{FA2}) immediately gives an upper bound for $g$:
\begin{eqnarray}
g < \sqrt{\frac{8\pi}{S}} \ .
\label{upper}
\end{eqnarray}
This is represented by the upper straight line of Fig.~\ref{unit} (a), for $S=0.1$ (top), $S=0.2$ (center), and $S=0.3$ (bottom). Additional upper and lower bounds are given by the requirement that the vector mesons are narrow states, since this is a necessary condition for the WSR's to hold. Above the upper left solid curve $M_{\rm V}$ is larger than $2M_{\rm A}$, and the dominant ${\rm V}\rightarrow {\rm A}+{\rm A}$ decay channel opens up, with a large contribution to $\Gamma_{\rm V}$. Below the lower thin solid curve $\Gamma_{\rm V}({\pi\pi})/M_{\rm V}$ is larger than 0.5. Here $\Gamma_{\rm V}({\pi\pi})$ is the partial width for decays into two $\pi$'s, which turns out to be
\begin{eqnarray}
\Gamma_{\rm V}({\pi\pi})=\frac{g^2_{V\pi\pi}}{96\pi}M_{\rm V}=\frac{M_{\rm V}^5}{96\pi g^2 F_\pi^4} \ .
\end{eqnarray}
Where in the last equation we used the relation $g_{V\pi\pi} =  M_{\rm V}^2/(g F^2_{\pi})$.  A final constraint is the requirement that $a$ is a positive number and ${\cal O}(1)$. On the lower thick curve $a=0$, and both WSR's are satisfied in a running regime. Below the curve $a$ is negative. Above the curve $a$ is positive and ${\cal O}(1)$, and all corresponding values of $M_{\rm A}$ and $g$ are possible.

We can use these results to analyze the $\pi-\pi$ scattering. We use nonlinear realizations in which the Higgs boson is integrated out. For the corresponding tree-level invariant amplitudes and scattering formalism see Appendix A of Ref. \cite{Sannino:1997fv} or directly equations (1) and (2) in \cite{Sannino:2007yp} after having set to zero the mass of the pions.  In Fig.~\ref{MA1TeV} (a) we plot the $I=0$, $J=0$ partial wave, $a_0^0$, for $M_{\rm A}$=1.5 TeV and three different values of $g$. The central curve has an $a_0^0=0.5$ maximum, and displays the largest value of unitarity violation energy, around 3.7 TeV, for $M_{\rm A}$=1.5 TeV. The lowest curve has a slightly lower value of $g$, and violates unitarity just below 3.7 TeV. However the upper curve violates unitarity at a much lower energy, around 1.9 TeV. This shows that there is a discontinuity in the plot of unitarity violation energy as a function of $g$, as shown in Fig.~\ref{MA1TeV} (b). The location of this discontinuity corresponds therefore to a ``critical'' value of $g$, $g_{\rm c}$. While above $g_{\rm c}$ the amplitude is not sufficiently unitarized by the vector-vector meson, and a spin-0 isospin-0 state is required with a pole mass near the energy scale where unitarity is violated \cite{Harada:2003em},  below and near $g_{\rm c}$ the theory may very well be Higgsless. Moreover, since $M_{\rm A}<M_{\rm V}$ for $M_{\rm A}=$1.5 TeV and $S=0.1$ (see Fig.~\ref{spectrum}), these states are more difficult to be discovered at the LHC or ILC reach. Of course this argument does not exclude the presence of light states, but shows clearly that a worst case scenario of no detection of new particles cannot be excluded as it is generally expected in models with a weak EWSB sector.

 Notice that for small values of $g$ the theory seems to lose unitarity rather soon, even below the standard 1.2 TeV bound of the chiral Lagrangian. This seemingly unreasonable result comes from holding $M_{\rm A}$ fixed, in which case taking small values of $g$ corresponds to taking large values of $g_{{\rm V}\pi\pi}$. In this limit, and for relatively small values of $M_{\rm V}$, the interaction between the vector-vector meson and the pions becomes too strong, and the model quickly looses unitarity. Another way to see this comes from Eq.~(\ref{FV}): taking the $g\rightarrow 0$ limit with $M_{\rm V}$ fixed gives $F_{\rm V}\rightarrow\infty$. It would be more physical to let $M_{\rm V}$ rapidly approach zero at the same time, in such a way that $F_{\rm V}\rightarrow 0$ as well. However here we only focus on the phenomenologically relevant regions of the parameter space, in which the vector mesons are expected to be at least in the $10^2$ GeV range.

In Fig.~\ref{unit} (a) the dashed curve gives $g_{\rm c}$ as a function of $M_{\rm A}$. Below and in the vicinity of this curve unitarity violation is delayed to higher energy scales, and a spin-0 isospin-0 state is not needed with mass below 1 TeV. Notice that for $S=0.1$ (top) and $S=0.2$ (center) $g_{\rm c}$ is only attained in the running regime, i.e. $a=0$,  when $\Gamma_{\rm V}/M_{\rm V}>0.5$, while for $S=0.3$ one still has $\Gamma_{\rm V}/M_{\rm V}<0.5$. This is in agreement with the results of Ref.~\cite{Harada:2003em,Sannino:2007yp}, and shows that QCD-like theories are unitarized by the vector meson only for large values of $S$. In Fig.~\ref{unit} (b) the unitarity violation energy for $g=g_{\rm c}$ is plotted as a function of $M_{\rm A}$. To the right of the vertical thin line $\Gamma_{\rm V}/M_{\rm V}$ becomes greater than 0.5. To the right of the vertical thick line $a$ becomes negative. Notice how the thin line is first hit by the dashed curve for $S=0.1$ and $S=0.2$, while the thick line is hit first for $S=0.3$.

 We analyzed the longitudinal scattering of the WW gauge bosons in scenarios of dynamical breaking of the electroweak theory of walking technicolor type. Due to the equivalence theorem this scattering is mapped directly into the pion-pion scattering of the underlying strongly coupled gauge theory driving the breaking of the electroweak theory. A similar analysis has been performed in \cite{Fabbrichesi:2007ad}. We have shown that there are regions in the parameters space of these models, allowed by the electroweak precision data, according to which tree-level unitarity is delayed up to around 3-4 TeV without the inclusion of any sub-TeV resonances.  We have also studied the case in which the axial vector state is heavier than the associated vector meson and found that in this case the vector meson alone cannot unitarize the scattering amplitude. 

\acknowledgments
We would like to thank D.D. Dietrich, M.T. Frandsen and T. Ryttov for useful discussions. Our work  is supported by the Marie Curie Excellence Grant under contract MEXT-CT-2004-013510.

\end{document}